\documentclass[12pt]{article}
\usepackage{graphicx}
\usepackage[round, authoryear]{natbib}
\usepackage{url} 

\usepackage{subfig}
\usepackage{mathrsfs,amsmath,amsthm,amssymb,color}
\usepackage{hyperref}
\usepackage{cleveref}
\usepackage[ruled,linesnumbered]{algorithm2e}
\usepackage{booktabs}
\usepackage{CJK}
\allowdisplaybreaks[4]
\def\D{\mathrm{d}}
\def\tr{\mbox{tr}}
\newtheorem{theorem}{Theorem}
\newtheorem{lemma}{Lemma}

\newcommand{\blind}{1}

\usepackage[pagewise]{lineno} 
\begin{document}

\def\spacingset#1{\renewcommand{\baselinestretch}%
{#1}\small\normalsize} \spacingset{1}

\date{}

\if0\blind
{
  \title{\bf Doubly Smoothed Density Estimation with Application on Miners' Unsafe Act Detection}
  
  \maketitle
} \fi

\if1\blind
{
  \title{\bf Doubly Smoothed Density Estimation with Application on Miners' Unsafe Act Detection}
  \author{Qianhan Zeng\hspace{.2cm}\\
    Guanghua School of Management, Peking University\\
    Miao Han \\
    School of Mathematics, China University of Mining and Technology\\
    Ke Xu \\
    School of Statistics, University of International Business and Economics\\
    Feifei Wang\footnote{Corresponding author: feifei.wang@ruc.edu.cn} \\
    Center for Applied Statistics and School of Statistics, Renmin University of China\\
    and \\
    Hansheng Wang \\
    Guanghua School of Management, Peking University}
  \maketitle
} \fi

\bigskip
\begin{abstract}
We study anomaly detection in images under a fixed-camera environment and propose a \emph{doubly smoothed} (DS) density estimator that exploits spatial structure to improve estimation accuracy.
The DS estimator applies kernel smoothing twice: first over the value domain to obtain location-wise classical nonparametric density (CD) estimates, and then over the spatial domain to borrow information from neighboring locations. 
Under appropriate regularity conditions, we show that the DS estimator achieves smaller asymptotic bias, variance, and mean squared error than the CD estimator. 
To address the increased computational cost of the DS estimator, we introduce a grid point approximation (GPA) technique that reduces the computation cost of inference without sacrificing the estimation accuracy. 
A rule-of-thumb bandwidth is derived for practical use. 
Extensive simulations show that GPA-DS achieves the lowest MSE with near real-time speed.
In a large-scale case study on underground mine surveillance, GPA-DS enables remarkable sub-image extraction of anomalous regions after which a lightweight MobileNet classifier achieves $\approx$99\% out-of-sample accuracy for unsafe act detection. 
\end{abstract}

\noindent%
{\it Keywords:}  Anomaly Detection; Convolutional Neural Networks; Deep Learning; Grid Point Approximation; Kernel Density Estimation; Nonparametric Kernel Smoothing.
\vfill

\newpage
\spacingset{1.5} 
\section{Introduction}
\label{sec:intro}

Anomaly detection in images is a problem that arises in many different settings.
In the industrial domain, anomaly detection has been widely used to identify defects in manufacturing processes \citep{patchcore_2022, industryAD_2023}.
In the medical domain, anomaly detection has become a key tool for identifying subtle disease patterns in magnetic resonance imaging, computed tomography, and pathology images \citep{medicalAD_2021}.
In the surveillance and safety domain, anomaly detection has been employed to detect abnormal human activities for effective safety monitoring and security \citep{sultani2018real}.
These diverse applications underscore the importance of developing both accurate and efficient anomaly detection methods.

Detecting anomalies in images has been studied from both supervised and unsupervised perspectives.
Supervised methods, often based on deep learning, are powerful but rely heavily on large labeled datasets and are limited to the anomaly types in the training labels \citep{medicalAD_2021}.
In contrast, unsupervised methods do not require labels and therefore offer greater flexibility in scenarios with limited or unavailable annotations.
Among these, two prominent directions are Robust Principal Component Analysis (RPCA)-based methods and density-based methods.
RPCA-based methods decompose data into background and foreground anomalies \citep{RPCA_2011}.
A typical drawback of such methods is their reliance on iterative matrix operations, which makes these methods computationally expensive \citep{RPCA_2011, SSSR_2018, JMLR_2022, JMLR_2024}.
Density-based methods estimate probability distributions and flag anomalies in low-density regions \citep{lowdensity_2020}.
However, they face two key limitations, with vulnerability to misspecified parametric assumptions \citep{OA_2017} and high computational costs in nonparametric settings \citep{lowdensity_2020}.
More importantly, they often neglect spatial contextual information, which might degrade estimation accuracy in image applications.
These limitations motivate the development of methods that both exploit spatial information and improve computational efficiency.

\begin{figure}[htbp]
	\centering
	\subfloat[Carrying oversized objects]{\label{subfig:a}\includegraphics[scale=0.07]{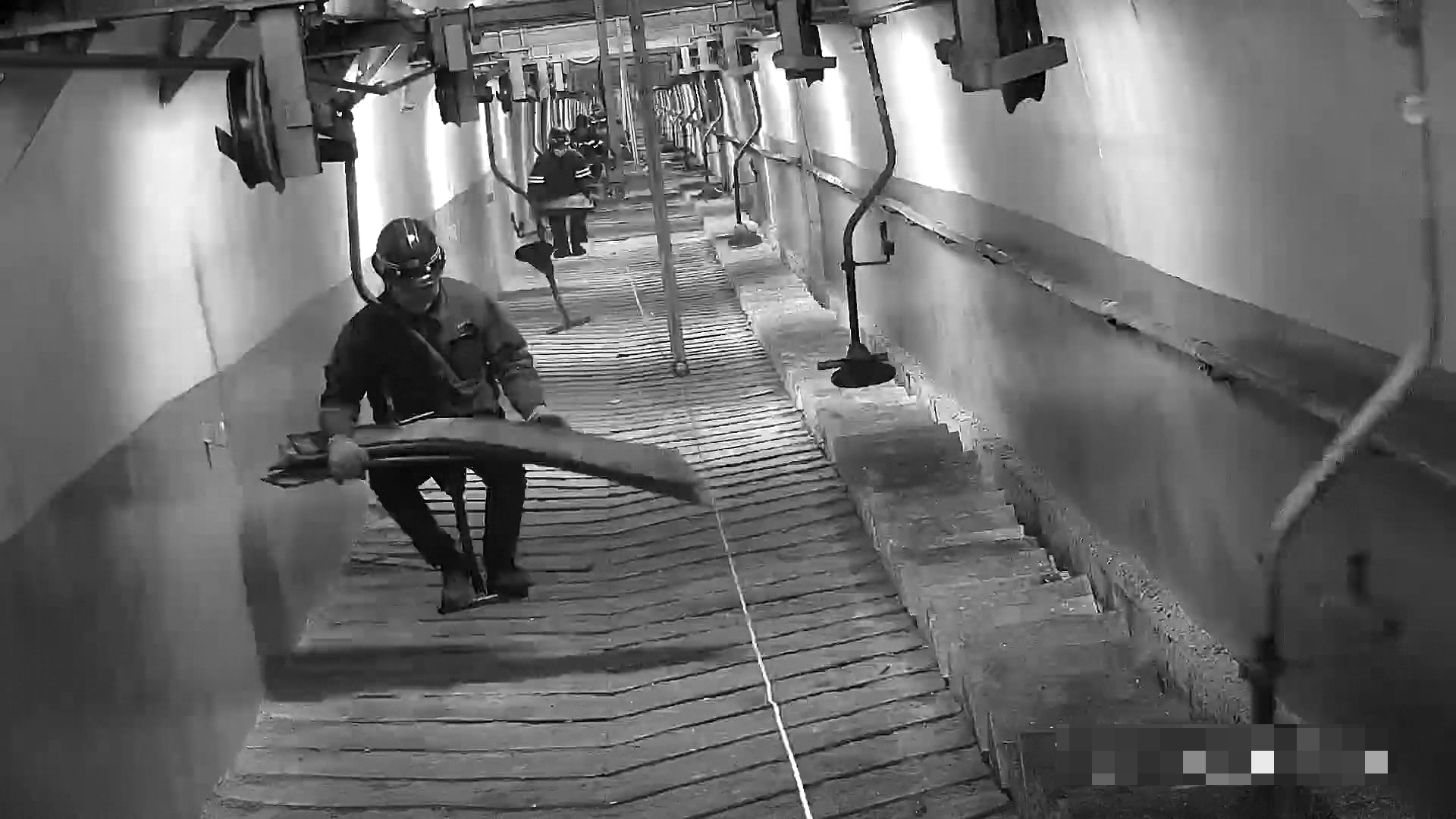}}\quad
	\subfloat[Dragging feet on the floor]{\label{subfig:b}\includegraphics[scale=0.07]{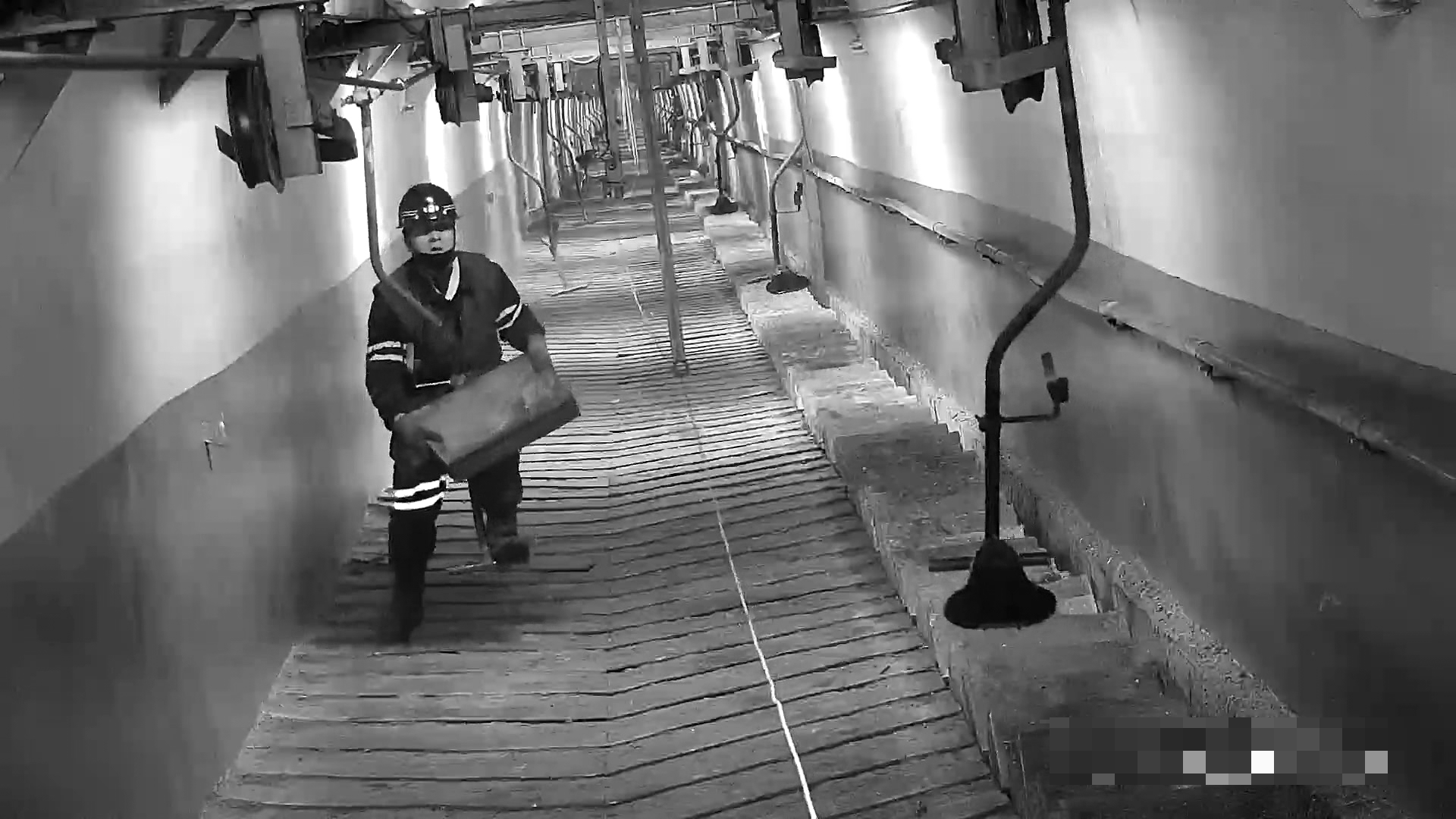}}\quad
	\subfloat[Walking miners]{\label{subfig:c}\includegraphics[scale=0.07]{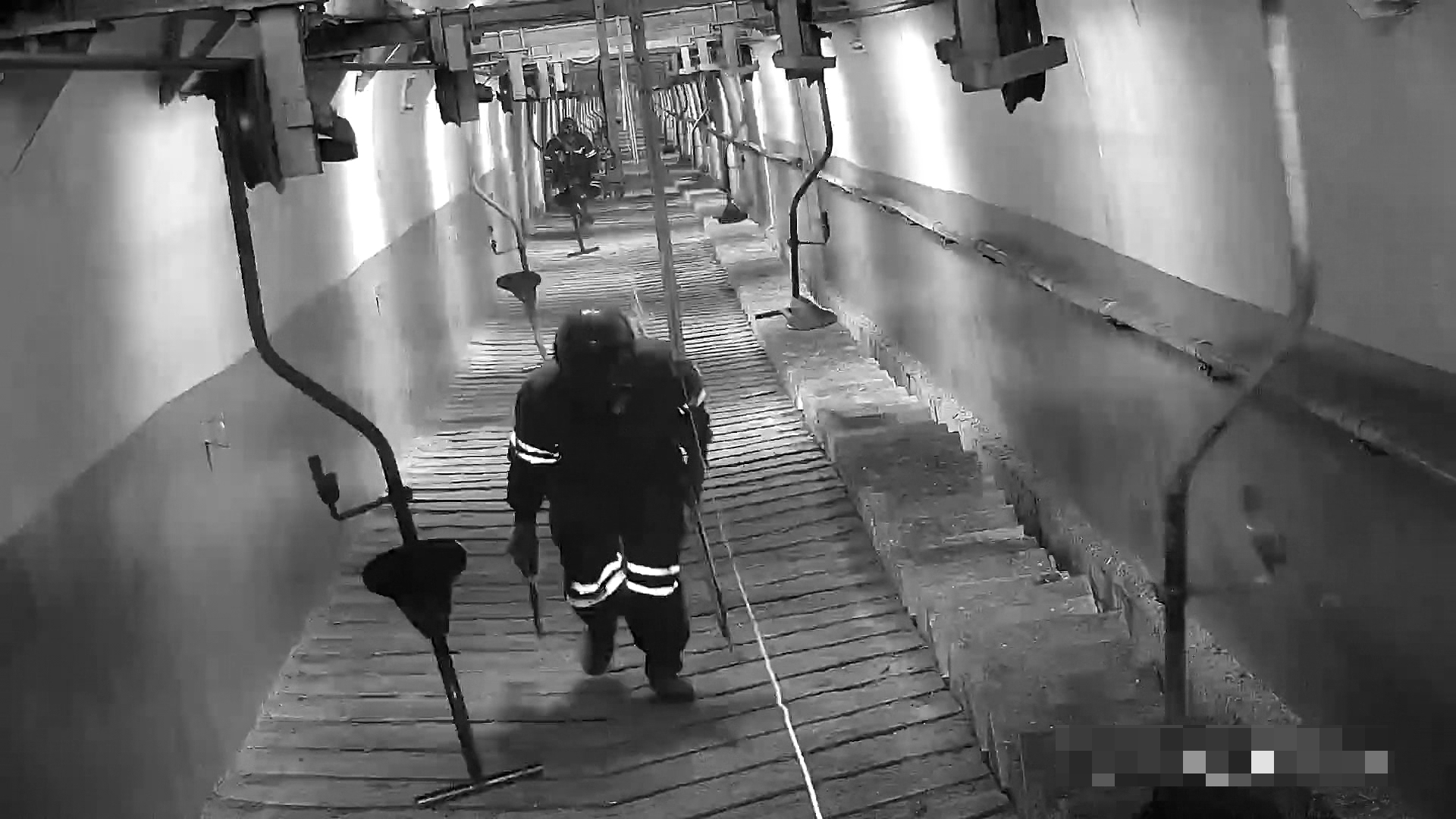}}
	\\	
	\caption{A motivating example in a fixed-camera environment. 
    The goal is to first localize miners and then identify potential unsafe acts. 
    Because the camera is stationary, each pixel corresponds to a consistent physical location in the real world, 
    allowing miners to be detected as anomalies in images. 
    Panels (a)--(c) illustrate examples of unsafe acts.
    }
	\label{fig: unsafe acts}
\end{figure}

As a motivating example, we consider the task of localizing miners and identifying potential unsafe acts in images captured by fixed surveillance cameras in underground mining environments; see Figure~\ref{fig: unsafe acts} for an example.
This application is particularly well-suited to density-based anomaly detection because each pixel consistently corresponds to a fixed physical location in the real world.
Moreover, due to the inherent spatial structure of imaging data, neighboring pixels often share similar background patterns and anomalies.
However, as noted earlier, existing density-based methods typically fail to fully exploit such valuable spatial information, which leads to suboptimal estimation and classification accuracy in practice.
Motivated by this consideration, we develop a novel doubly smoothed (DS) density estimator, which integrates valuable spatial information from neighboring pixel locations through an additional kernel smoothing over the spatial domain.
Under appropriate regularity conditions, we are able to show rigorously that the asymptotic bias, variance, and mean squared error (MSE) of the DS estimator are of smaller order than those of the classical nonparametric density (CD) estimator.
Although the DS estimator achieves improved theoretical properties, its computational cost substantially increases due to the additional kernel smoothing.
To overcome this issue, we introduce a grid point approximation (GPA) technique that accelerates computation without compromising estimation accuracy. 
Extensive numerical experiments demonstrate that the DS estimator, combined with the GPA method and rule-of-thumb bandwidths, delivers excellent empirical performance.

The remainder of this article is organized as follows.
Section~\ref{sec:literature} provides a brief review of existing anomaly detection methods.
Section~\ref{sec:Density Estimation} introduces the proposed DS estimator, establishes its statistical properties, and presents the GPA method for improving computational efficiency.
In Section~\ref{sec:Application of Unsafe Act Detection}, we report extensive numerical experiments demonstrating the superiority of the GPA-DS estimator, followed by a case study on detecting miners' unsafe acts. 
Section~\ref{sec:conclusion} concludes the article with a brief discussion.

\section{Literature Review} \label{sec:literature}

In this section, we briefly review existing research on anomaly detection in images. 
Existing approaches can be roughly categorized into supervised and unsupervised (or semi-supervised) methods.
Supervised methods typically exploit deep learning-based object detection models, while unsupervised methods rely on learning the representation of normal patterns to identify deviations.

Supervised methods have achieved remarkable progress with the advent of convolutional neural networks (CNNs), which form the backbone of many influential models such as R-CNN \citep{RCNN_2014}, Fast R-CNN \citep{FastRCNN_2015}, Faster R-CNN \citep{FasterRCNN_2015}, and YOLO \citep{Yolo_2016, yolov8_2023}.
These methods are praised for their generality and flexibility, allowing detection across diverse and complex backgrounds.
However, the resulting models are inherently limited to the abnormality types represented in the training data, and their performance heavily depends on the availability of extensive and diverse annotated data \citep{medicalAD_2021}.
Annotating such data typically requires both bounding boxes and class labels, and bounding box annotation is far more labor-intensive and costly than classification labeling.
\citet{onebit_2020} reported that binary image-level labeling takes roughly 2.7 seconds per image, whereas annotation of a single bounding box takes a median of 42 seconds \citep{lff_2012}. 
In terms of financial cost, workers on Amazon Mechanical Turk charge bounding box annotation approximately three times higher than that of classification labeling \citep{kili2025annotationpricing}.

To address the challenges of supervised methods, many unsupervised or semi-supervised methods have been developed.
For example, \cite{tech_2018} proposed a spatio-temporal smooth sparse decomposition that separates data into functional mean, sparse anomalies, and random noise.
More recently, \cite{patchcore_2022} introduced PatchCore, which leverages ImageNet pre-trained embeddings for anomaly detection.
It extracts features from nominal image patches during training and then detects anomalies during testing by comparing new patches against this reference set.
Another influential line of work is the Robust Principal Component Analysis (RPCA), which decomposes a data matrix into a sparse foreground and a low-rank background \citep{RPCA_2011}. 
Extensions of RPCA have addressed mixed pixels \citep{JMLR_2022}, data heterogeneity \citep{JMLR_2024}, spatiotemporal foreground structures \citep{SSSR_2018}, and non-linear features \citep{RDA_2017}.
While conceptually intuitive, RPCA-based methods often suffer from high computational costs due to iterative singular value decomposition and related operations \citep{RPCA_2011, SSSR_2018, JMLR_2022, JMLR_2024}.
To reduce computational complexity, \cite{LRPCA_2021} proposed LRPCA using deep unfolding for computation acceleration.

Another important line of unsupervised anomaly detection is density-based methods, which can be broadly divided into parametric and nonparametric approaches.
Parametric methods assume a specific form for the data distributions, such as the Gaussian distribution \citep{3sigma_2000, gaussian_2014} or mixtures of parametric forms \citep{mixture_2000, GMMreview_2018}.
While these models are simple and computationally efficient, their performance can deteriorate significantly when the assumed distribution deviates from the true data-generating process \citep{OA_2017}.
In contrast, nonparametric methods make minimal assumptions about the underlying distribution and offer greater flexibility.
Among them, kernel density estimation (KDE) has been widely applied in video anomaly detection and background modeling, where each pixel is modeled by a kernel density estimator \citep{KDE_2002, Li2014An, local_2020}.
However, KDE-based methods often suffer from high computational complexity, especially when processing high-resolution images.
Moreover, their statistical efficiency is limited because they do not fully exploit the spatial information among neighboring pixels.

\section{Density Estimation}
\label{sec:Density Estimation}

Density estimation provides a principled foundation for anomaly detection by modeling the underlying data distribution.
Instead of relying on explicit labels, density-based anomaly detection methods estimate the probability density and then identify anomalies as locations with low density.
For image data, this task requires accurately and efficiently estimating the density at each spatial location.

In this section, we first introduce the CD estimator in Section~\ref{sec: The CD Estimator}, which performs kernel smoothing over the value domain at each fixed location.
In Section~\ref{sec: The DS Estimator}, we then extend this framework by incorporating an additional layer of kernel smoothing across the spatial domain, leading to the proposed DS estimator.
The theoretical properties of the DS estimator are studied in Section \ref{sec: Theoretical Properties}, followed by the development of a practical rule-of-thumb bandwidth selection strategy in Section \ref{sec: Rule-of-Thumb Bandwidth}.
In Section~\ref{sec: Grid Point Approximation}, we introduce a grid point approximation method to substantially reduce the computational burden while retaining the desirable asymptotic properties of the DS estimator.

\subsection{The CD Estimator} \label{sec: The CD Estimator}

Suppose there are $N$ images provided.
Each image is assumed to have been appropriately rescaled to lie within a spatial domain $\mathbb{D} = [0, 1] \times [0, 1]$. 
Let $X_i = \{X_i(s): s=(s_{(1)}, s_{(2)})^\top \in \mathbb{D} \}$ denote the $i$th image for $1 \leq i \leq N$. 
At any location $s \in \mathbb{D}$, we treat $X_i(s)$ as a random variable defined on the value domain $\mathbb{V} = [0,1]$ with probability density function $f(x, s)$.
The density function satisfies $f(x, s) = 0$ for all $(x, s) \notin \mathbb{V} \times \mathbb{D}$ and $f(x, s) \geq 0$ for any $(x, s) \in \mathbb{V} \times \mathbb{D}$. 
Accordingly, the density is normalized such that $\int_0^1 f(x,s) \D{x}=1$ for all $s \in \mathbb{D}$. 
Different $X_i(s)$s are assumed to be mutually independent. 
Given a location $s \in \mathbb{D}$, a classic nonparametric density estimator for $f(x, s)$ can be constructed as
\begin{align}
    \widehat{f}_{\text{CD}}(x, s) = \left(\frac{1}{N h}\right) \sum_{i=1}^N K \! \left(\frac{ X_i(s) - x}{h} \right).
    \label{eq: CD}
\end{align}
For convenience, we refer to \eqref{eq: CD} as the CD estimator.
Here, $K(t)$ is a kernel function, and $h > 0$ denotes the bandwidth.
Typically, $K(t)$ is chosen as a bounded density function that is symmetric about zero (see, e.g., equation (1.10) in \cite{nonparametric_2007}).

Computationally, the CD estimator applies kernel smoothing to the $N$ pixel values $\{X_i(s): 1 \leq i \leq N\}$ at a given location $s$, where the kernel smoothing is essentially a kernel-weighted average. 
As a result, the computational complexity of evaluating the CD estimator for a single pair $(x, s)$ is $O(N)$.
Theoretically, the asymptotic bias, variance, and MSE of the CD estimator in \eqref{eq: CD} are well-established for fixed locations, as demonstrated in \cite{Fan_1996}, \cite{pagan_ullah_1999}, and \cite{nonparametric_2007}.
For ease of reference, we restate Theorem~1.1 of \cite{nonparametric_2007} as Lemma~\ref{lemma:KDE}, along with the necessary notations.
Specifically, let $\mu_m = \int t^m K(t) \D t$ and $\nu_m = \int t^m K^2(t) \D t$ for $m\geq0$.
Denote the first-order derivative of $f(x, s)$ with respect to $x$ by $\dot{f}_x(x, s)=\partial f(x, s)/\partial x \in \mathbb{R}$ and the gradient with respect to $s$ by $\dot{f}_s(x,s)= \left\{ \partial f(x, s)/\partial s_{(1)}, \partial f(x, s)/\partial s_{(2)} \right\}^\top \in \mathbb{R}^2$.
Similarly, $\ddot{f}_x(x, s)=\partial^2 f(x, s)/\partial x^2 \in \mathbb{R}$ denotes the second-order derivative of $f(x, s)$ with respect to $x$, and $\ddot{f}_s(x,s)$ denotes the $2 \times 2$ Hessian matrix of $f(x,s)$ with respect to $s$.

\begin{lemma} (Theorem~1.1 of \cite{nonparametric_2007}) \label{lemma:KDE}
    At a fixed location $s$, assume that $X_1(s), \cdots, X_N(s)$ are independently and identically distributed observations having a three-times differentiable probability density function $f(x, s)$ over $x$.
    Let $x$ be an interior point in $\mathbb{V}$ and the kernel function $K(t)$ is a bounded density function symmetric about zero with $\mu_2 > 0$.
    Also, as $N \rightarrow \infty$, $h \rightarrow 0$ and $N h \rightarrow \infty$.
    Then, we have
    \begin{gather}
        E \left\{ \widehat{f}_\mathrm{CD}(x, s) \right\} = f(x, s) + \left( \frac{\mu_2 h^2 }{2} \right) \ddot{f}_x(x, s) + o(h^2), \nonumber \\
        \mathrm{var} \left\{ \widehat{f}_\mathrm{CD}(x, s) \right\} = \left( \frac{1}{N h} \right) \left\{ \nu_0 f(x, s) + o(1) \right\}. \nonumber 
    \end{gather}
\end{lemma}

\subsection{The DS Estimator} \label{sec: The DS Estimator}

As one can see, the CD estimator in \eqref{eq: CD} smooths over the value domain $\mathbb{V}$ by giving higher weights to pixel values $X_i(s)$ that lie closer to the target value $x$.
A natural extension is to apply an additional layer of smoothing to the CD estimates across the spatial domain $\mathbb{D}$, giving greater weight to estimates from locations closer to the target $s$.
This extension is motivated by the observation that nearby locations often share similar distributional patterns.
Therefore, borrowing information from nearby estimates may improve density estimation.

As a first attempt, we define the spatially smoothed version of $\widehat{f}_{\text{CD}}(x,s)$ as
\[
\widehat{f}^*(x, s) = h^{-2} \int_\mathbb{D} \widehat{f}_\text{CD}(x, v) \mathbb{K} \{ ( v - s ) / h \} \D v,
\]
where $\mathbb{K}(s) = K(s_{(1)}) K(s_{(2)})$ denotes a two-dimensional kernel function.
This construction is conceptually straightforward.
It replaces each estimate $\widehat{f}_{\text{CD}}(x,s)$ with a weighted average of estimates from nearby locations, where the weight is determined by the kernel function $\mathbb{K}(\cdot)$ and the bandwidth $h$.
In principle, the kernel function $K(\cdot)$ and the bandwidth $h$ could differ from those used in \eqref{eq: CD}.
However, we set them to be identical for simplicity.

Unfortunately, the computation of $\widehat{f}^*(x, s)$ is infeasible in practice, as it would require evaluating $\widehat{f}_{\text{CD}}(x, s)$ at every location in $\mathbb{D}$.
Instead, observations are available only at a finite set of locations.
Let $\mathbb{S}_M = \{s_m : 1 \leq m \leq M\} \subset \mathbb{D}$ denote the set of observed locations, with cardinality $|\mathbb{S}_M| = M$.
For theoretical development, these locations $s_m$ are assumed to be independently generated from a uniform distribution.
(An extension to evenly spaced locations is provided in Appendix F of the supplementary materials.)
Based on the observed locations, we then construct a feasible estimator as
\begin{gather}
    \widehat{f}_{\text{DS}}(x, s) = \sum_{m=1}^M \widehat{f}_{\text{CD}}(x, s_m) \omega_m, \label{eq:DS}
\end{gather}
where the weight is given by $\omega_m = \mathbb{K}\{ (s_m - s) / h \} / \sum_{m=1}^M \mathbb{K}\{ (s_m - s) / h \}$ so that $\sum_{m=1}^M \omega_m = 1$.
For convenience, we refer to \eqref{eq:DS} as the DS estimator, since it applies a second layer of kernel smoothing over the spatial domain $\mathbb{D}$.
Computationally, the DS estimator averages over $M$ pre-computed CD estimators $\{ \widehat{f}_\text{CD}(x, s_m) : 1 \leq m \leq M \}$.
As a result, the computational complexity of evaluating the DS estimator for a single pair $(x, s)$ is $O(NM)$, which is substantially more expensive than the $O(N)$ cost of the CD estimator.
In the next section, we analyze the asymptotic properties of the DS estimator to assess the benefit of this additional spatial smoothing.

\subsection{Theoretical Properties} \label{sec: Theoretical Properties}

To study the theoretical properties of the DS estimator $\widehat{f}_{\text{DS}}(x,s)$, the following technical conditions are needed.
\begin{itemize}
\item[(C1)] (\textit{Kernel Function}) The kernel function $K(t)$ is a symmetric probability density function with $\int |t|^m K(t) \D t < \infty$ and $\int |t|^m K^2(t) \D t <\infty$ for $m \leq 3$.
\item[(C2)] (\textit{Density Function}) Given any interior location $s \in \mathbb{D}$, the probability density function $f(x, s)$ is sufficiently smooth so that its third-order partial derivatives with respect to both $x$ and $s$ are continuous.
\item[(C3)] (\textit{Bandwidths and Sample Size}) Assume $h \rightarrow 0$, $N h \rightarrow \infty$, and $M / N^\alpha \rightarrow C_M$ for some constants $\alpha > 0$ and $C_M > 0$ as $N \rightarrow \infty$.
\end{itemize}
By Condition (C1), we assume that the kernel function $K(t)$ is a well-behaved probability density function, which should be symmetric about zero. By Condition (C2), the probability function $f(x, s)$ should be sufficiently smooth in both $x$ and $s$.
For a given $s$, $f(x, s)$ represents the probability density function in $x$.
The requirement for the third-order partial derivatives of $f(x, s)$ with respect to $x$ aligns with the conditions presented in \cite{nonparametric_2007}. For a given $x$, Condition (C2) also requires the probability function $f(x, s)$ with respect to $s$ to be smooth. 
By Condition (C3), we require that the bandwidth $h$ should converge to zero at an appropriate speed as the sample size goes to infinity.
In particular, we require the locally effective sample size to be sufficiently large, that is $Nh \rightarrow \infty$.
{In practice, both the number of observed locations $M$ and sample size $N$ are typically large.}
Therefore, it is theoretically appealing to consider the asymptotic regime, where $M \rightarrow \infty$ as $N \rightarrow \infty$.
To this end, the relative diverging rate between $M$ and $N$ must be analytically specified.
Condition (C3) basically imposes such a specification in the form of $M = C_M N^\alpha$.
Accordingly, both $C_M$ and $\alpha$ should be regarded as fixed constants rather than tunable parameters.
Except for the Condition (C3), which requires $M/N^\alpha \rightarrow C_M$ for some fixed constant $C_M > 0$, all assumptions made in (C1)--(C3) are very standard conditions, which have been widely used in the literature \citep{silverman_1986, Fan_1996, pagan_ullah_1999, nonparametric_2007}.
Next, we study the theoretical properties of the DS estimator $\widehat{f}_{\text{DS}}(x,s)$ in the following Theorem~\ref{Theorem 1}, whose proof can be found in Appendix A of the supplementary materials.

\begin{theorem} \label{Theorem 1}
Assume conditions (C1)--(C3), and $Nh^3 \rightarrow 0$ as $N \rightarrow \infty$. Then, we have
\begin{gather}
    E\Big\{\widehat{f}_{{\rm DS}}(x, s)\Big\} = f(x,s) + \left(\frac{\mu_2h^2}{2}\right)\ddot{f}_x(x, s) + \left(\frac{\mu_2 h^2}{2}\right) {\rm tr}\Big\{\ddot{f}_s(x, s)\Big\} + o\big(h^2\big), \nonumber \\
    {\rm var}\Big\{\widehat{f}_{{\rm DS}}(x, s)\Big\} = \left(\frac{1}{NM h^3}\right) \Big\{ \nu_0^3 f(x, s) + o(1) \Big\}. \nonumber
\end{gather}
\end{theorem}

Next, we compare the asymptotic bias, variance, and MSE between the CD and DS estimators under their corresponding optimal bandwidths.
First, the MSE of the CD estimator is of order $O(h^4 + N^{-1} h^{-1})$ based on Lemma~\ref{lemma:KDE}.
Minimizing the leading terms of the MSE with respect to the bandwidth $h$ yields the optimal bandwidth
$h_{\text{CD}} = C_\text{CD} N^{-1/5}$ for some bandwidth constant $C_{\text{CD}} > 0$.
Substituting this back gives $\mathrm{bias} \{ \widehat{f}_\mathrm{CD}(x, s) \} = O(N^{-2/5})$, $\mathrm{var}\{ \widehat{f}_\mathrm{CD}(x, s) \} = O(N^{-4/5})$, and $\mathrm{MSE} \{ \widehat{f}_{\text{CD}}(x, s) \} = O(N^{-4/5})$.
Second, the MSE of the DS estimator is of order $O(h^4 + N^{-1} M^{-1} h^{-3})$ based on Theorem~\ref{Theorem 1}.
Minimizing the leading terms of the MSE with respect to the bandwidth $h$ yields the optimal bandwidth $h_\text{DS} = C_\text{DS} N^{-1/7} M^{-1/7}$ for some bandwidth constant $C_{\text{DS}} > 0$.
Substituting this back gives $\mathrm{bias}\{ \widehat{f}_\mathrm{DS}(x, s) \} = O(N^{-2/7} M^{-2/7})$, $\mathrm{var}\{ \widehat{f}_\mathrm{DS}(x, s) \} = O(N^{-4/7} M^{-4/7})$, and $\mathrm{MSE}\{ \widehat{f}_\mathrm{DS}(x, s) \} = O(N^{-4/7} M^{-4/7})$.
Then, under the condition that $M \geq N^{2/5}$, the asymptotic bias, variance, and MSE of the DS estimator have a smaller order than those of the CD estimator.

\subsection{Rule-of-Thumb Bandwidth} \label{sec: Rule-of-Thumb Bandwidth}

We are then motivated to determine the bandwidth constant $C_\text{DS}$.
The leading term of the MSE of $\widehat{f}_{\text{DS}}(x, s)$ can be expressed as $\mathcal{L}_{\text{MSE}}^{(x,s)}\left(h \right) = C_1(x,s) / (NMh^3) + C_2(x,s) h^4$, where $C_1(x,s) = \nu_0^3 f(x,s)$ and $C_2(x,s)=\mu_2^2 [ \ddot{f}_x(x, s) + \tr\{\ddot{f}_s(x,s)\} ]^2 / 4$, with $\mathrm{tr}(\cdot)$ denoting the trace operator.
To account for all possible values of $x$ and $s$, we approximate the integrated MSE as in \cite{pagan_ullah_1999}: $\mathcal{L}_{\text{MSE}}(h) = \int_0^1 \int_{\mathbb{D}} \mathcal{L}_{\text{MSE}}^{(x,s)}(h) \D{x}\D{s} = C_1 / \left(NMh^3\right) + C_2 h^4$, where $C_1 = \int_0^1 \int_{\mathbb{D}} C_1(x,s) \D{x}\D{s}$ and $C_2 = \int_0^1 \int_{\mathbb{D}} C_2(x,s) \D{x}\D{s}$. 
Minimizing $\mathcal{L}_{\text{MSE}}(h)$ with respect to $h$ leads to the optimal bandwidth constant as $C_\text{DS} = \left\{3C_1 / (4C_2) \right\}^{1/7}$.
Consequently, we have $\mathcal{L}_{\text{MSE}}(h_{\text{opt}})= \{\left(4/3\right)^{3/7}+\left(3/4\right)^{4/7}\} C_1^{4/7} C_2^{3/7}  (N M)^{-4/7}$.

In practice, computing the optimal bandwidth remains a challenging task, primarily because the key constants $C_1$ and $C_2$ are unknown.
A widely used approach for bandwidth selection is cross-validation or generalized cross-validation \citep{GCV_1985, Fan_1996, nonparametric_2007}.
However, in our setting, both the sample size $N$ and the number of observed locations $M$ are large, making the computational cost of cross-validation prohibitively high.
As a practical alternative, we adopt a simpler yet effective solution by following the rule-of-thumb approach proposed by \cite{silverman_1986}.

The basic idea is to adopt a working form for the density function $f(x,s)$ that enables analytical evaluation of the unknown constants $C_1$ and $C_2$.
Specifically, we assume that $f(x,s) = \exp\{-(x-0.5)^2/2\sigma^2\}/ (c\sqrt{2\pi}\sigma )$, corresponding to the probability density function of a truncated normal distribution.
Here, $c = 1 - 2\Phi(-0.5/\sigma)$ and $\Phi (\cdot )$ denotes the cumulative distribution function of the standard normal distribution.
When $\sigma$ is small, the truncation effect near the boundaries is negligible, and $c$ can be approximated by $1$.
This approximation is well justified in our empirical setting, where the estimated standard deviation from the MineLaneway dataset is approximately $\sigma = 0.16$.
Next, assume a standard Gaussian kernel function is used. 
We then have
\begin{align}
    C_1 =& \int_0^1 \int_{\mathbb{D}} C_1(x,s) \D{x}\D{s} = \int_0^1 \int_{\mathbb{D}} {\nu_0^3 f(x,s)} \D{x}\D{s}
    = \nu_0^3 = \frac{1}{8}\pi^{-\frac{3}{2}}. \nonumber
\end{align} 
Note that $\ddot{f}_x(x,s)=\left\{(x-0.5)^2 / \sigma^2 - 1\right\} e^{-(x-0.5)^2/2\sigma^2}/\left(\sqrt{2\pi}\sigma^3\right)$, $\dot{f}_s(x,s) = \left(0, 0\right)^\top$, and $\ddot{f}_s(x,s) = \left(\bf{0}\right)_{2\times2}$. We then have
\begin{align}
    C_2 = \!\int_0^1\!\! \int_{\mathbb{D}} C_2(x,s) \D{x}\D{s} = \left( \frac{\mu_2^2}{4} \right) \!\int_0^1\! \int_{\mathbb{D}} \left[\ddot{f}_x(x,s) + \tr\left\{\ddot{f}_s(x,s)\right\} \right]^2 \D{x} \D{s} \approx \frac{3}{32} \pi^{-1/2} \sigma^{-5}. \nonumber
\end{align} 
This leads to the bandwidth constant $C_{\text{DS}} = \pi^{-1/7} \sigma^{5/7}$. 
For practical implementation, $\sigma$ can be approximated by the empirical standard deviation of the sample. 
Our extensive numerical studies suggest that this simple rule-of-thumb method performs remarkably well in practice.

\subsection{Grid Point Approximation} \label{sec: Grid Point Approximation}

Although the DS estimator provides improved estimation accuracy, its computational cost remains high.
Evaluating $\widehat{f}_{\text{DS}}(\overline{x}, \overline{s})$ for a single observed pair $(\overline{x}, \overline{s})$ requires $O(NM)$ operations.
To address this challenge, we introduce a GPA method that substantially reduces the computational burden without compromising estimation accuracy.
The key idea of GPA is to employ a pre-computation strategy during the training phase, allowing for rapid density evaluation at the testing phase.

In the training phase, we pre-compute the DS estimators $f(x_g^*, \overline{s})$ at each grid point $x_g^*$ for every location of interest $\overline{s} \in \mathbb{S}_M$.
The grid points $x_g^*$ are randomly generated from the value domain $\mathbb{V}$. 
Let $\mathbb{G}^*=\{ x_g^*: x_g^* \in \mathbb{V}, 1 \leq g \leq G^*\}$ denote the set of all grid points, where $|\mathbb{G}^*| = G^{*}$.
We aim to keep $G^{*}$ substantially smaller than $NM$; otherwise, the computational savings would be negligible.
In the testing phase, we utilize the pre-computed DS estimators $\{\widehat{f}_{\text{DS}}(x_g^*, \overline{s}): x_g^* \in \mathbb{G}^*, \overline{s} \in \mathbb{S}_M \}$ to construct a density estimator for a given observed pair $(\overline{x}, \overline{s})$.
Specifically, we compute a weighted average of the pre-computed estimators as
\begin{gather}
    \widehat{f}_\text{GPA} (\overline{x},\overline{s}) = \sum_{g =1}^{G^*} \widehat{f}_\text{DS}(x_g^*, \overline{s}) \omega_x(x_g^*, \overline{x}), \label{eq:GPA-DS}
\end{gather}
where the weight is given by $\omega_x(x_g^*, \overline{x}) = K\{ (x_g^* - \overline{x}) / h^* \} / \sum_{g=1}^{G^*} K\{ (x_g^* - \overline{x}) / h^* \}$.
Here, $h^*$ is another bandwidth, which needs to be carefully selected.
One possible specification is $\alpha = 3/2$, $h = C_h N^{-5/14}$, $h^* = N^{-5/7}$, $G^* = N^{5/7} \log{N}/2$.
It can be verified that $h^* / h \rightarrow 0$, $Nh^3 = o(1)$, $G^*h^{*} \rightarrow \infty$, and $h^* N^{4(1+\alpha)/7} / G^* = o(1)$ as $N \rightarrow \infty$. 
This specification will be later justified by the conditions in the following Theorem~\ref{Theorem 2}.

We then refer to \eqref{eq:GPA-DS} as the GPA-DS estimator for $f(\overline{x}, \overline{s})$.
Compared with $\widehat{f}_{\text{DS}}(\overline{x}, \overline{s})$, one obvious advantage of $\widehat{f}_{\text{GPA}}(\overline{x}, \overline{s})$ is that, the computational complexity is greatly reduced from $O(NM)$ to $O(G^{*})$.
This makes real-time anomaly detection practically feasible.
The algorithmic details of the GPA-DS method for object detection can be found in Appendix C of the supplementary materials. 
In Appendix C, we also provide a detailed comparison between GPA-DS and ST-SSD \citep{tech_2018}, SRTC \citep{JMLR_2022}, and PerPCA \citep{JMLR_2024}, highlighting the computational advantages of GPA-DS during inference.
The theoretical properties of $\widehat{f}_{\text{GPA}}(\overline{x}, \overline{s})$ are given in Theorem \ref{Theorem 2}, whose proof can be found in Appendix B in supplementary materials.

\begin{theorem} \label{Theorem 2}
Assume conditions (C1)--(C3). Further assume that there exists a positive constant $\alpha > 4/3$, such that $h = C_h N^{-(1+\alpha)/7}$ for some $C_h > 0$, $h^* N^{(1+\alpha)/7} \rightarrow 0$, $G^*h^{*} \rightarrow \infty$, $ h^{*} N^{4(1+\alpha)/7} / G^* = o(1)$ as $N \rightarrow \infty$. Then, we have
\[
    E\Big\{\widehat{f}_{\rm GPA}\big(\overline{x}, \overline{s}\big)\Big\} = f\Big(\overline{x}, \overline{s}\Big) + \left(\frac{\mu_2h^2}{2}\right)\ddot{f}_x\Big(\overline{x},\overline{s}\Big)+ \left(\frac{\mu_2h^2}{2}\right){\rm tr }\Big\{\ddot{f}_s\Big(\overline{x},\overline{s}\Big)\Big\} + o\big(h^2\big),\]
\[{\rm var}\Big\{\widehat{f}_{{\rm GPA}}\big(\overline{x}, \overline{s}\big)\Big\} = \left( \frac{1}{NMh^3}\right) \Big\{\nu_0^3 f(\overline{x}, \overline{s}) + o(1)\Big\}.\]
\end{theorem}
\noindent 
Comparing the results of $\widehat{f}_{\text{GPA}}(\overline{x}, \overline{s})$ in Theorem \ref{Theorem 2} with those $\widehat{f}_{\text{DS}}(\overline{x}, \overline{s})$ in Theorem \ref{Theorem 1}, we find that the GPA-DS estimator shares the same leading terms in both asymptotic bias and asymptotic variance as the DS estimator.
At the same time, the computational cost of $\widehat{f}_{\text{GPA}}(\overline{x}, \overline{s})$ is substantially reduced.

\section{Numerical Studies}
\label{sec:Application of Unsafe Act Detection}

In this section, we first introduce in Section~\ref{subsec:dataset} a large-scale surveillance dataset collected from an underground mining environment. 
The unique characteristics of this dataset make it particularly suitable for demonstrating our methodology.
Various summary statistics (e.g., mean and variance) are computed from this dataset and then used for the following simulation specification in Section~\ref{sec: simulation}.
In Section~\ref{sec: simulation}, we conduct extensive simulation studies to compare the finite-sample performance of the proposed methods against various alternatives.
The results highlight the effectiveness of the GPA-DS method in density estimation.
In Section~\ref{sec:case_study}, we apply deep learning models to classify unsafe acts using sub-images extracted by the GPA-DS method.

\subsection{The MineLaneway Dataset}
\label{subsec:dataset}

We first give a brief introduction to the MineLaneway dataset to be demonstrated subsequently for anomaly detection in images.
This dataset is collected by extracting images from a nearly two-hour surveillance video recorded in a Chinese mining company.
The dataset contains a total of 90,549 high-resolution images of size $540 \times 960 \times 3$ (see Figure~\ref{fig: unsafe acts} for example images). 
The primary purpose of this surveillance system is to monitor miners' behaviors as they pass through a mine laneway. 
The mine laneway under study is a semi-enclosed corridor for miners to enter or exit the underground mines, which is usually dark, narrow, and risky.
Monkey cars, commonly used in China, are efficient vehicles that transport miners between the surface and underground areas.

The MineLaneway dataset further exhibits several unique characteristics that offer promising opportunities for anomaly detection by our DS and/or GPA-DS methods. 
First, all images are captured by a surveillance camera with a fixed position and angle, ensuring that every pixel has a consistent physical meaning. 
Consequently, pixels at the same spatial location across different images are directly comparable. 
Second, the pixel values at a fixed location typically show minimal variability over time, except when anomalies occur. 
This spatiotemporal stability allows us to identify objects of interest as anomalies against a nearly invariant background.

\subsection{Simulation Study}
\label{sec: simulation}

\begin{figure}[h]
    \centering
    \includegraphics[scale=0.7]{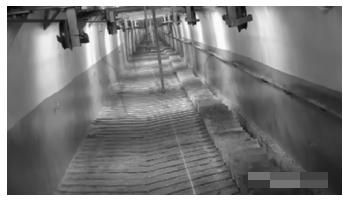}
    \caption{The graphical display of $\mu(s)$.}
    \vskip -0.1in
    \label{fig:mu}
\end{figure}

We next present a simulation study designed to examine the finite-sample performance of the proposed method.
Although the MineLaneway dataset consists of three-channel RGB images, all images are converted to a single-channel grayscale format for simplicity of analysis.
Let $X_i(s) \in \mathbb{V}$ be the $i$th simulated image with $1 \leq i \leq N$.
For every $s \in \mathbb{D}$, $X_i(s)$ is assumed to follow a truncated normal distribution with mean $\mu(s)$ and standard deviation $\sigma$.
The values of $\mu(s)$ are set to be the empirical smoothed counterparts as calculated on the aforementioned MineLaneway dataset, which is graphically displayed in Figure \ref{fig:mu}.
The value of $\sigma$ is set to be 0.16, which is roughly the same as the overall standard deviation of the MineLaneway dataset. 
Assume that $X_i(s)$ is observed on a finite number of locations with coordinates given by $\mathbb{S}_M = \{(s_{(1)}/p, s_{(2)}/q):1\leq s_{(1)} \leq p, 1 \leq s_{(2)} \leq q\}$ with $p = 540$ and $q = 960$.

To compute the GPA-DS estimator, we first compute $\widehat{f}_\text{DS}(x_g^*, s_m)$ for every $s_m \in \mathbb{S}_M$ at a grid point set $\mathbb{G}^* = \{ x_g^*: 1 \leq g \leq G^* \}$, where each $x_g^*$ is randomly generated on $\mathbb{V}$ from a uniform distribution.
Here, we set the number of the grid points as $G^* = 500$ and the kernel function as the standard Gaussian kernel.
According to the rule-of-thumb method, the bandwidth $h$ is set to $h = 0.013$ with $N=1{,}000$, and the bandwidth $h^*$ is specified as $h^*=5 h^2$.
Each image $X_i(s)$ has dimensions of $p = 540$ and $q = 960$, yielding an inter-pixel spacing of $\min\{1/p, 1/q\}$.
This spacing is extremely small, which is approximately 8\% of the bandwidth ($h = 0.013$).
This indicates that the spatial grid is much finer than the effective smoothing scale.
Consequently, the kernel weights assigned by the Gaussian kernel to location $s_m=(s_{m(1)}, s_{m(2)})^\top \in \mathbb{D}$ with $\max\{|s_{(1)} - s_{m(1)}|, |s_{(2)} - s_{m(2)}|\} > 3$ can be practically regarded as zero.
This truncation significantly reduces computational cost and can be efficiently implemented as a standard convolution operation in deep learning frameworks such as TensorFlow.

\begin{figure}[ht]
    \centering
    \vskip -0.1in
    \includegraphics[scale=0.6]{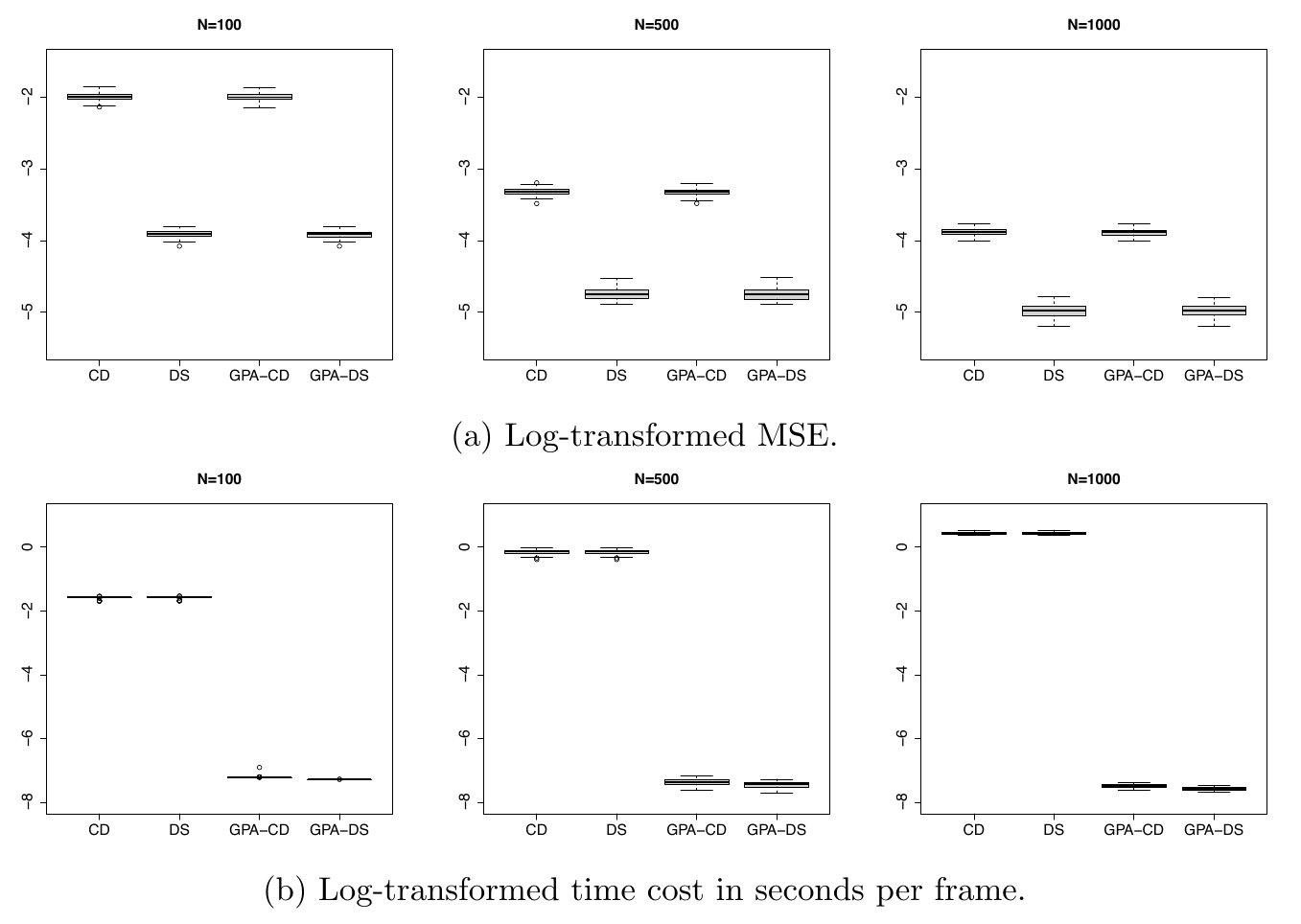}
    \vskip 0.1in
    \caption{The log-transformed MSE values and the log-transformed time costs per frame of the CD, DS, GPA-CD, and GPA-DS estimators with different sample sizes. }
    \vskip -0.1in
    \label{fig:MSE}
\end{figure}

To evaluate the estimation accuracy of the GPA-DS estimator, we randomly generate another set of test points from a uniform distribution as $x_g^+ \in \mathbb{V}$ with $1 \leq g \leq G^+=100$.
Then, the estimation accuracy of the GPA-DS estimator in one replication can be measured by the MSE as
\begin{gather*}
    \text{MSE}\Big(\widehat{f}_{\text{GPA}}\, \Big) = \left(\frac{1}{G^+|\mathbb{S}_M|}\right) \sum_{g=1}^{G^+} \sum_{s \in \mathbb{S}_M} \bigg\{\widehat{f}_{\text{GPA}}\Big(x_{g}^+, s\Big) - f\Big(x_{g}^+, s\Big) \bigg\}^2.
\end{gather*}
For a reliable evaluation, this experiment is randomly replicated for a total of $T=100$ times.
This leads to a total of $T$ MSE values, which are then log-transformed and boxplotted in Figure~\ref{fig:MSE}(a).
For comparison purposes, the MSE values of the CD estimator and the DS estimator are also evaluated.
Furthermore, we also include the MSE values of the GPA-CD estimator for comparison, where the GPA-CD estimator is computed by applying the GPA method to the CD estimator based on the grid points $\mathbb{G}^*$.

By Figure~\ref{fig:MSE}(a), we can draw the following conclusions.
First, the MSE values of the CD, DS, GPA-CD, and GPA-DS estimators steadily decrease as the sample size increases, yielding more accurate density estimates.
Second, the CD and GPA-CD estimators exhibit inferior MSE performance because they rely solely on location-wise information without incorporating spatial information.
In contrast, the DS and GPA-DS estimators achieve substantially lower MSE by borrowing information from neighboring locations.
Lastly, the DS and GPA-DS estimators exhibit comparable MSE performance.
This corroborates the theoretical result established in Theorem~\ref{Theorem 2}.
The key difference lies in computational efficiency.
The average computation time per frame (in seconds) for each method is reported in Figure~\ref{fig:MSE}(b).
Our simulation results show that the GPA-DS and GPA-CD estimators require approximately 0.0006 seconds on average to compute the density estimate for a single testing image.
This remarkable efficiency enables both estimators to perform real-time density estimation.
In contrast, the DS and CD estimators are considerably more computationally demanding, taking approximately 1.5 seconds to process a single image when $N = 1{,}000$.
In summary, the GPA-DS method has the best MSE performance and computational efficiency compared with other alternatives.

\subsection{Case Study}
\label{sec:case_study}

Building on the density estimation framework developed in the previous sections, we now apply the proposed method to a case study on detecting unsafe acts of miners using surveillance images from the MineLaneway dataset.
The case study involves two primary tasks: localizing miners and classifying their behaviors as safe or unsafe.
To accomplish the first task, we employ the GPA-DS estimator to identify anomalous regions and extract them as sub-images, as described in Section~\ref{sec:Sub-Image Extraction}.
The empirical performance and computational efficiency of this GPA-DS-based sub-image extraction method are evaluated and compared with alternative methods.
Section~\ref{sec:Deep Learning-based Classification} then presents a deep learning-based classification model trained on the extracted sub-images, which are manually annotated to indicate the presence or absence of unsafe acts.

\subsubsection{Sub-Image Extraction}
\label{sec:Sub-Image Extraction}

\begin{figure}[htbp]
	\centering
	\subfloat[An example image.]{\label{subfig-example}\includegraphics[scale=0.6]{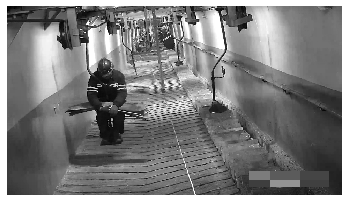}}\quad\quad
	\subfloat[Rescaled density map by the GPA-DS estimator.]{\label{subfig-GPA_density}\includegraphics[scale=0.5]{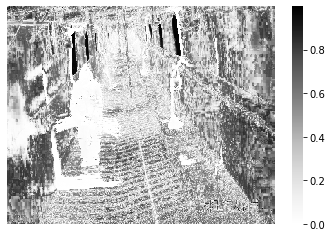}}
	\\	
	\caption{(a) Example image showing a miner taking a monkey car and passing through the laneway; (b) estimated density map by the GPA-DS estimator (rescaled to [0, 1]).}
	\label{fig:suspected}
\end{figure}

We first develop a GPA-DS-based procedure to extract sub-images that correspond to anomalous regions identified in the surveillance images.
We randomly select 2,000 images from the MineLaneway dataset to compute $\widehat{f}_\text{DS}(x_g^*, s_m)$ for any grid point $x_g^* \in \mathbb{G}^*$ and any observed location $s_m \in \mathbb{S}_M$.
The number of grid points is set to $G^* = 500$.
Given the pre-computed $\{ \widehat{f}_\text{DS}(x_g^*, s_m): x_g^* \in \mathbb{G}^*, s_m \in \mathbb{S}_M \}$, we then compute the GPA-DS estimator according to \eqref{eq:GPA-DS} with $\overline{x}$ as the observed pixel value at location $\overline{s} = s_m$.
To provide an intuitive illustration, Figure~\ref{fig:suspected}(a) shows an example image in which a miner is taking a monkey car in the left area.
Figure~\ref{fig:suspected}(b) presents the corresponding GPA-DS density map (rescaled to the range $[0, 1]$).
The estimated density clearly highlights an anomalous region in the same left area, corresponding to the moving miner.


\begin{figure}[htbp]
	\centering
	\subfloat[Remove background.]{\label{subfig:1}\includegraphics[scale=0.41]{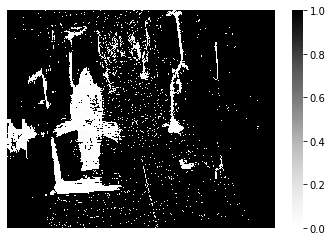}}\quad
	\subfloat[Blurred image.]{\label{subfig:2}\includegraphics[scale=0.41]{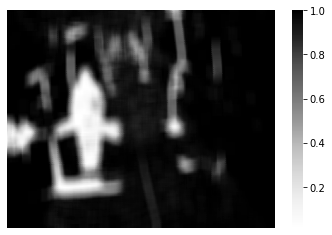}}\quad
	\subfloat[Binary image.]{\label{subfig:3}\includegraphics[scale=0.4]{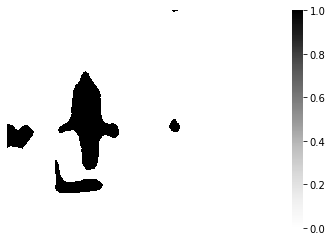}}
	\\	
	\caption{(a) is the outcome after removing the background of the density map. (b) shows the blurred image. (c) is the outcome after binarizing the blurred image.}
	\label{fig:remove}
\end{figure}

From an implementation perspective, the following procedure is adopted to accurately locate the anomalous region.
First, in Figure~\ref{fig:suspected}(b), background pixels with values greater than the threshold $\alpha_1 = 0.06$ are removed, yielding the image shown in Figure~\ref{fig:remove}(a).
Second, the resulting image is smoothed using a $33 \times 33$ average pooling filter to produce more spatially coherent regions, as illustrated in Figure~\ref{fig:remove}(b).
Third, the pixel values are binarized as the value of 1 if smaller than $\alpha_2 = 0.42$ and 0 otherwise.
This results in the binary map shown in Figure~\ref{fig:remove}(c).
Lastly, connected components with areas smaller than 5,500 pixels are discarded.
Among the remaining regions, the one with the largest area is extracted as the sub-image.
This completes the GPA-DS-based sub-image extraction method.
It is worth noting that the GPA-DS estimator can be replaced by alternative density estimators (e.g., CD, DS, or GPA-CD), leading to other density-based sub-image extraction variants.


\begin{figure}[htbp]
	\centering
	\vskip -0.2in
	\subfloat[]{\label{subfig-box}\includegraphics[scale=0.51]{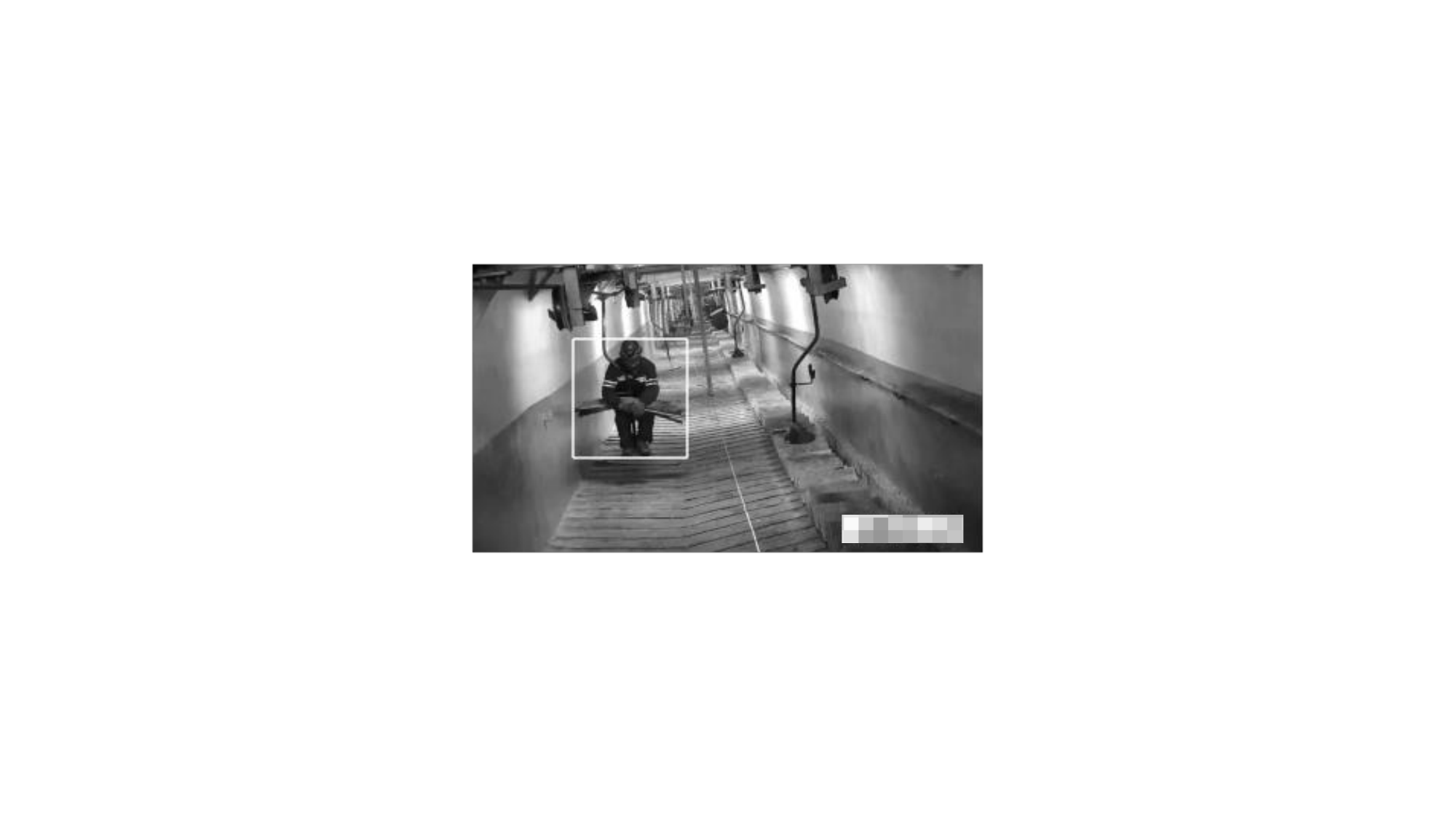}}\quad
	\subfloat[]{\label{subfig-bar}\includegraphics[scale=0.29]{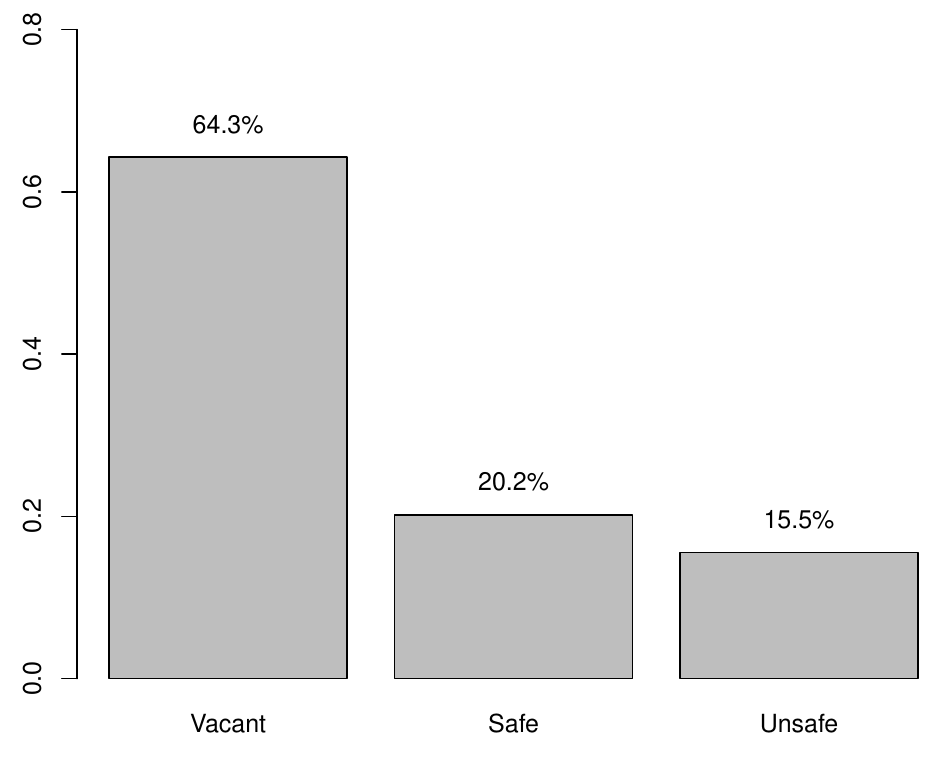}}
	\\	
	\caption{(a) Example of an unsafe act highlighted by the ground-truth bounding box; (b) a barplot showing the percentage distribution of class labels.}
	\label{fig: bounding box}
\end{figure}

Next, we evaluate the empirical performance of the proposed sub-image extraction method.
Although the proposed method is unsupervised, ground-truth bounding box annotations are still required for performance assessment.
Ideally, tight bounding box annotation should be provided for every image; see the white box in Figure \ref{fig: bounding box}(a) for illustrations.
However, manually annotating all $N=90{,}549$ images would be prohibitively time-consuming and labor-intensive.
To reduce the annotation burden, only a subset of the images is labeled.
Specifically, the subset consists of the first frame captured every second of the surveillance video, which contains 25 frames per second.
This leads to a total of $90{,}549 / 25 \approx 3{,}622$ sampled images, which are subsequently annotated by field experts with both class labels and bounding boxes.
There are three types of labels: (1) \textit{vacant}, when no miners are present; (2) \textit{safe}, when miners are performing normal activities; and (3) \textit{unsafe}, when miners are engaging in unsafe acts.
The proportions of each class are shown in Figure~\ref{fig: bounding box}(b).

Next, we evaluate the finite-sample performance of the proposed sub-image extraction method.
For a comprehensive comparison, a number of unsupervised and supervised methods are also included.
Those methods are, respectively, the RPCA method \citep{RPCA_2011}, the LRPCA method \citep{LRPCA_2021}, the B-SSSR method \citep{SSSR_2018}, and the YOLOv8 model \citep{yolov8_2023}.
For the YOLOv8 model, both the pre-trained and fine-tuned versions are considered.
The details of the model implementation are given in Appendix D of the supplementary materials.

We then evaluate the aforementioned methods from three perspectives.
First, we measure the average computational time per frame (in seconds) required for bounding box prediction.
Second, we compute the image-level F1 score \citep{NIPS_2014}, defined as $\text{F}1 = 2 \times \text{TP} / (2 \times \text{TP} + \text{FP} + \text{FN})$.
Here, TP is the number of true positives, FN is the number of false negatives, and FP is the number of false positives.
A positive case is defined as a case where a bounding box is detected for an image.
A negative case is defined as a case where no bounding box is detected for an image.
Third, we calculate the Intersection-of-Union (IoU) \citep{NEURIPS2021_a8f15eda}, defined as $\text{IoU} = |R_1 \cap R_2| / |R_1 \cup R_2|$.
Here, $R_1$ is the region of the annotated bounding box, $R_2$ is the region of the predicted bounding box, and $|R|$ stands for the number of pixel points in an arbitrary region $R$.
The detailed comparison results are summarized in Table~\ref{table: comparison}.
Representative predicted bounding boxes produced by the methods listed in Table~\ref{table: comparison} are presented in Appendix E of the supplementary materials.

\begin{table}[htbp]
\small
\centering
\renewcommand\arraystretch{0.8}
\caption{Comparison results for the methods considered on the evaluation set of the MineLaneway dataset, with the best results highlighted in \textbf{bold}.
Here, ``Avg" represents the average, ``Med" denotes the median, ``U" represents unsupervised, and ``S" denotes supervised.
}
\begin{tabular}{cccccc}
\hline
Method             & Type & Avg Time       & Avg F1         & Avg IoU        & Med IoU        \\ \hline
GPA-DS             & U    & 0.033          & \textbf{0.955} & 0.770          & 0.835          \\
GPA-CD             & U    & 0.036          & 0.942          & 0.756          & 0.827          \\
DS                 & U    & 2.540          & 0.947          & \textbf{0.771} & 0.838          \\
CD                 & U    & 2.250          & 0.941          & 0.755          & 0.826          \\
RPCA               & U    & 0.259          & 0.644          & 0.564          & 0.647          \\
LRPCA              & U    & \textbf{0.029} & 0.639          & 0.536          & 0.615          \\
B-SSSR             & U    & 0.970          & 0.704          & 0.599          & 0.712          \\
PatchCore          & U    & 0.055           & 0.772          & 0.532          & 0.676          \\
Pre-trained YOLOv8 & U/S  & 0.031          & 0.767          & 0.588          & 0.715          \\
Fine-tuned YOLOv8  & S    & 0.031          & 0.941          & 0.747          & \textbf{0.906} \\ \hline
\end{tabular}
\label{table: comparison}
\end{table}

We can draw the following important conclusions from Table~\ref{table: comparison}.
First, both the GPA-DS-based sub-image extraction method and the fine-tuned YOLOv8 model \citep{yolov8_2023} demonstrate the best overall performances in terms of both computational time and prediction accuracy (i.e., average F1 score, average IoU, and median IoU).
However, the key difference is that no bounding box information is required for the GPA-DS-based method.
In contrast, bounding box information is necessarily needed for the YOLOv8 model.
Second, the overall performances of the GPA-DS-based method seem to be the best among all the unsupervised methods in terms of both computational time and prediction accuracy.
Next, the GPA-DS-based method demonstrates prediction accuracy similar to the DS-based method.
They both outperform their CD counterparts clearly.
In particular, the GPA-DS-based method outperforms the GPA-CD-based method materially by a margin of 1.3\% in average F1 score, 1.4\% in average IoU, and 0.8\% in median IoU.
Similar patterns are also observed for non-GPA methods (i.e., DS- and CD-based methods).
However, the GPA-CD- and GPA-DS-based methods are significantly more computationally efficient than their non-GPA counterparts.

\subsubsection{Unsafe Act Detection}
\label{sec:Deep Learning-based Classification}

We next turn to detecting unsafe acts using the sub-images extracted by the GPA-DS-based method.
Specifically, we apply the GPA-DS-based sub-image extraction method to each of the ten consecutive images in the MineLaneway dataset.
Let $X_{i}^{\text{sub}}(s)$ be the potential sub-image extracted from the $i$th image with $i \in \{10, 20, \cdots, \lfloor N / 10 \rfloor \times 10\} $.
Each sub-image is appropriately rescaled to the spatial domain $\mathbb{D}$, with observed locations $\mathbb{S}^{\text{sub}} = \{ (s_{(1)} / r, s_{(2)} / r): 1 \leq s_{(1)} \leq r, 1 \leq s_{(2)} \leq r \}$ and $r = 200$.
Each extracted sub-image $X_i^{\text{sub}}(s)$ is manually annotated by field experts with a class label $Y_i$ indicating vacant ($Y_{i}=-1$), safe ($Y_{i}=0$), or unsafe ($Y_{i}=1$).
This annotation process results in a total of 3,428 labeled sub-images, which serve as the dataset for subsequent model training and evaluation.

The annotated sub-images are randomly divided into a training set (80\%) and a testing set (20\%).
A CNN model is used to extract high-dimensional features for each sub-image.
Specifically, we adopt the MobileNet architecture \citep{MobileNet}, which comprises 86 layers and approximately 3.23 million parameters pre-trained on the ImageNet dataset.
Each sub-image is encoded into a 50,176-dimensional feature vector, which is subsequently used to train a multinomial logistic regression classifier on the training set.


Model performance is evaluated on the testing set in terms of prediction accuracy.
To ensure robustness, the entire process is repeated 100 times, with the data randomly re-partitioned in each replication.
This leads to a total of 100 prediction accuracy values.
The mean and median prediction accuracy are 99.45\% and 99.41\%, respectively.
This indicates remarkable performance of classifying miners' unsafe acts.
For comparison, we also train the same MobileNet model directly on the full images with image-level annotations.
After 100 replications, this baseline achieves a mean prediction accuracy of 90.09\% and a median accuracy of 90.03\%.
These results clearly demonstrate that the sub-image-based method outperforms the full-image baseline, likely because sub-images are easier to classify due to reduced visual complexity and clearer class-specific features.

\section{Conclusion}
\label{sec:conclusion}

In this work, we propose a density-based method for anomaly detection in images.
Motivated by the spatial structure of images, we develop the DS density estimator, which incorporates spatial information from neighboring pixel locations to improve estimation accuracy.
To further improve computational efficiency, we propose the GPA technique.
Both rigorous theory and extensive numerical experiments demonstrate the superiority of the GPA-DS method.
In the case study, the GPA-DS method is shown to be effective in localizing miners as anomalies. 
Their unsafe acts are then identified using deep learning models with outstanding prediction accuracy.

To conclude this article, we consider some interesting topics for potential future studies. 
As a density estimation method that incorporates spatial information, the proposed GPA-DS estimator can be extended to other applications where spatial structure plays a critical role.
For example, in environmental studies, estimating the density of pollutant concentrations across locations is important for monitoring air quality.
Nearby locations often exhibit similar pollutant distributions due to similar meteorological and topographical conditions.
As another example, in urban studies, estimating crime density is important for public safety.
Criminal activities tend to follow similar patterns in neighboring areas rather than occur uniformly across space.
Exploring these extensions of the GPA-DS method constitutes an interesting direction for future research.

 \section*{Acknowledgments}
 Qianhan Zeng sincerely thanks both Peking University and the University of Michigan for their academic support and inspiring research environment. 
 Most of the research and analysis were conducted at Peking University, and the manuscript was subsequently restructured and refined at the University of Michigan.
 This research is supported by National Natural Science Foundation of China (No.72371241, 72495123, 12271012), the MOE Project of Key Research Institute of Humanities and Social Sciences (22JJD910002), and the Big Data and Responsible Artificial Intelligence for National Governance, Renmin University of China. 
 
 \section*{Conflict of Interest}
 The authors report there are no competing interests to declare.







\bibliographystyle{asa}

\bibliography{Bibliography-MM-MC}

\begin{thebibliography}{36}
\newcommand{\enquote}[1]{``#1''}
\expandafter\ifx\csname natexlab\endcsname\relax\def\natexlab#1{#1}\fi

\bibitem[{Aggarwal(2017)}]{OA_2017}
Aggarwal, C.~C. (2017), \textit{An Introduction to Outlier Analysis}, Cham: Springer International Publishing, pp. 1--34.

\bibitem[{Baur et~al.(2021)Baur, Denner, Wiestler, Navab, and Albarqouni}]{medicalAD_2021}
Baur, C., Denner, S., Wiestler, B., Navab, N., and Albarqouni, S. (2021), \enquote{Autoencoders for Unsupervised Anomaly Segmentation in Brain MR Images: A Comparative Study,} \textit{Medical Image Analysis}, 69, 101952.

\bibitem[{Cai et~al.(2021)Cai, Liu, and Yin}]{LRPCA_2021}
Cai, H., Liu, J., and Yin, W. (2021), \enquote{Learned Robust PCA: A Scalable Deep Unfolding Approach for High-Dimensional Outlier Detection,} \textit{Advances in Neural Information Processing Systems}, 34, 16977--16989.

\bibitem[{Cand{\`e}s et~al.(2011)Cand{\`e}s, Li, Ma, and Wright}]{RPCA_2011}
Cand{\`e}s, E.~J., Li, X., Ma, Y., and Wright, J. (2011), \enquote{Robust Principal Component Analysis?} \textit{Journal of the ACM}, 58, 1--37.

\bibitem[{Elgammal et~al.(2002)Elgammal, Duraiswami, Harwood, and Davis}]{KDE_2002}
Elgammal, A., Duraiswami, R., Harwood, D., and Davis, L. (2002), \enquote{Background and Foreground Modeling Using Nonparametric Kernel Density Estimation for Visual Surveillance,} \textit{Proceedings of the IEEE}, 90, 1151--1163.

\bibitem[{Eskin(2000)}]{mixture_2000}
Eskin, E. (2000), \enquote{Anomaly Detection over Noisy Data using Learned Probability Distributions.} in \textit{Proceedings of the Seventeenth International Conference on Machine Learning}, pp. 255--262.

\bibitem[{Fan and Gijbels(1996)}]{Fan_1996}
Fan, J. and Gijbels, I. (1996), \textit{Local Polynomial Modelling and Its Applications: Monographs on Statistics and Applied Probability}, vol.~66, Chapman \& Hall/CRC.

\bibitem[{Girshick(2015)}]{FastRCNN_2015}
Girshick, R. (2015), \enquote{Fast R-CNN,} in \textit{Proceedings of the IEEE International Conference on Computer Vision}, pp. 1440--1448.

\bibitem[{Girshick et~al.(2014)Girshick, Donahue, Darrell, and Malik}]{RCNN_2014}
Girshick, R., Donahue, J., Darrell, T., and Malik, J. (2014), \enquote{Rich Feature Hierarchies for Accurate Object Detection and Semantic Segmentation,} in \textit{Proceedings of the IEEE Conference on Computer Vision and Pattern Recognition}, pp. 580--587.

\bibitem[{Goyal and Singhai(2018)}]{GMMreview_2018}
Goyal, K. and Singhai, J. (2018), \enquote{Review of Background Subtraction Methods Using Gaussian Mixture Model for Video Surveillance Systems,} \textit{Artificial Intelligence Review}, 50, 241--259.

\bibitem[{He et~al.(2021)He, Erfani, Ma, Bailey, Chi, and Hua}]{NEURIPS2021_a8f15eda}
He, J., Erfani, S., Ma, X., Bailey, J., Chi, Y., and Hua, X.-S. (2021), \enquote{$\alpha$-IoU: A Family of Power Intersection over Union Losses for Bounding Box Regression,} in \textit{Advances in Neural Information Processing Systems}, vol.~34, pp. 20230--20242.

\bibitem[{Howard et~al.(2017)Howard, Zhu, Chen, Kalenichenko, Wang, Weyand, Andreetto, and Adam}]{MobileNet}
Howard, A.~G., Zhu, M., Chen, B., Kalenichenko, D., Wang, W., Weyand, T., Andreetto, M., and Adam, H. (2017), \enquote{{MobileNets}: Efficient Convolutional Neural Networks for Mobile Vision Applications,} \textit{arXiv preprint arXiv:1704.04861}.

\bibitem[{Hu et~al.(2020{\natexlab{a}})Hu, Xie, Du, Hong, and Tian}]{onebit_2020}
Hu, H., Xie, L., Du, Z., Hong, R., and Tian, Q. (2020{\natexlab{a}}), \enquote{One-bit Supervision for Image Classification,} \textit{Advances in Neural Information Processing Systems}, 33, 501--511.

\bibitem[{Hu et~al.(2020{\natexlab{b}})Hu, Gao, Li, Wu, Du, and Maybank}]{local_2020}
Hu, W., Gao, J., Li, B., Wu, O., Du, J., and Maybank, S. (2020{\natexlab{b}}), \enquote{Anomaly Detection Using Local Kernel Density Estimation and Context-Based Regression,} \textit{IEEE Transactions on Knowledge and Data Engineering}, 32, 218--233.

\bibitem[{Javed et~al.(2019)Javed, Mahmood, Al-Maadeed, Bouwmans, and Jung}]{SSSR_2018}
Javed, S., Mahmood, A., Al-Maadeed, S., Bouwmans, T., and Jung, S.~K. (2019), \enquote{Moving Object Detection in Complex Scene Using Spatiotemporal Structured-Sparse RPCA,} \textit{IEEE Transactions on Image Processing}, 28, 1007--1022.

\bibitem[{{Kili Technology}(2025)}]{kili2025annotationpricing}
{Kili Technology} (2025), \enquote{Estimating Image Annotation Pricing for AI Projects,} Accessed: 2025-05-03.

\bibitem[{Li et~al.(2014)Li, Li, and Zhou}]{Li2014An}
Li, B., Li, Y., and Zhou, H. (2014), \enquote{An Improved Kernel Density Estimation Approach for Moving Objects Detection,} \textit{Open Automation and Control Systems Journal}, 6, 768--781.

\bibitem[{Li and Racine(2007)}]{nonparametric_2007}
Li, Q. and Racine, J.~S. (2007), \textit{Nonparametric Econometrics: Theory and Practice}, Princeton University Press.

\bibitem[{McKenna et~al.(2000)McKenna, Jabri, Duric, Rosenfeld, and Wechsler}]{3sigma_2000}
McKenna, S.~J., Jabri, S., Duric, Z., Rosenfeld, A., and Wechsler, H. (2000), \enquote{Tracking Groups of People,} \textit{Computer Vision and Image Understanding}, 80, 42--56.

\bibitem[{Pagan and Ullah(1999)}]{pagan_ullah_1999}
Pagan, A. and Ullah, A. (1999), \textit{Nonparametric Econometrics}, Themes in Modern Econometrics, Cambridge University Press.

\bibitem[{Puthiya~Parambath et~al.(2014)Puthiya~Parambath, Usunier, and Grandvalet}]{NIPS_2014}
Puthiya~Parambath, S., Usunier, N., and Grandvalet, Y. (2014), \enquote{Optimizing F-Measures by Cost-Sensitive Classification,} in \textit{Advances in Neural Information Processing Systems}, vol.~27.

\bibitem[{Redmon et~al.(2016)Redmon, Divvala, Girshick, and Farhadi}]{Yolo_2016}
Redmon, J., Divvala, S., Girshick, R., and Farhadi, A. (2016), \enquote{You Only Look Once: Unified, Real-Time Object Detection,} in \textit{Proceedings of the IEEE Conference on Computer Vision and Pattern Recognition}, pp. 779--788.

\bibitem[{Reis et~al.(2023)Reis, Kupec, Hong, and Daoudi}]{yolov8_2023}
Reis, D., Kupec, J., Hong, J., and Daoudi, A. (2023), \enquote{Real-Time Flying Object Detection with YOLOv8,} \textit{arXiv preprint arXiv:2305.09972}.

\bibitem[{Ren et~al.(2015)Ren, He, Girshick, and Sun}]{FasterRCNN_2015}
Ren, S., He, K., Girshick, R., and Sun, J. (2015), \enquote{Faster R-CNN: Towards Real-Time Object Detection with Region Proposal Networks,} in \textit{Advances in Neural Information Processing Systems}, eds. Cortes, C., Lawrence, N., Lee, D., Sugiyama, M., and Garnett, R., Curran Associates, Inc., vol.~28.

\bibitem[{Roth et~al.(2022)Roth, Pemula, Zepeda, Sch{\"o}lkopf, Brox, and Gehler}]{patchcore_2022}
Roth, K., Pemula, L., Zepeda, J., Sch{\"o}lkopf, B., Brox, T., and Gehler, P. (2022), \enquote{Towards Total Recall in Industrial Anomaly Detection,} in \textit{Proceedings of the IEEE Conference on Computer Vision and Pattern Recognition}, pp. 14318--14328.

\bibitem[{Shaikh and Kitagawa(2014)}]{gaussian_2014}
Shaikh, S.~A. and Kitagawa, H. (2014), \enquote{Efficient Distance-Based Outlier Detection on Uncertain Datasets of Gaussian Distribution,} \textit{World Wide Web}, 17, 511--538.

\bibitem[{Shen et~al.(2022)Shen, Xie, and Kong}]{JMLR_2022}
Shen, B., Xie, W., and Kong, Z.~J. (2022), \enquote{Smooth Robust Tensor Completion for Background/Foreground Separation with Missing Pixels: Novel Algorithm with Convergence Guarantee,} \textit{Journal of Machine Learning Research}, 23, 1--40.

\bibitem[{Shi and Kontar(2024)}]{JMLR_2024}
Shi, N. and Kontar, R.~A. (2024), \enquote{Personalized PCA: Decoupling Shared and Unique Features,} \textit{Journal of Machine Learning Research}, 25, 1--82.

\bibitem[{Silverman(1986)}]{silverman_1986}
Silverman, B.~W. (1986), \textit{Density Estimation for Statistics and Data Analysis}, Chapman \& Hall/CRC Taylor \& Francis Group.

\bibitem[{Su et~al.(2012)Su, Deng, and Fei-Fei}]{lff_2012}
Su, H., Deng, J., and Fei-Fei, L. (2012), \enquote{Crowdsourcing Annotations for Visual Object Detection,} in \textit{Proceedings of the Human Computation Workshop at the 26th AAAI Conference on Artificial Intelligence}, AAAI Workshop, pp. 40--46.

\bibitem[{Sultani et~al.(2018)Sultani, Chen, and Shah}]{sultani2018real}
Sultani, W., Chen, C., and Shah, M. (2018), \enquote{Real-World Anomaly Detection in Surveillance Videos,} in \textit{Proceedings of the IEEE Conference on Computer Vision and Pattern Recognition}, pp. 6479--6488.

\bibitem[{Thudumu et~al.(2020)Thudumu, Branch, Jin, and Singh}]{lowdensity_2020}
Thudumu, S., Branch, P., Jin, J., and Singh, J. (2020), \enquote{A Comprehensive Survey of Anomaly Detection Techniques for High Dimensional Big Data,} \textit{Journal of Big Data}, 7, 42.

\bibitem[{Wahba(1985)}]{GCV_1985}
Wahba, G. (1985), \enquote{{A Comparison of GCV and GML for Choosing the Smoothing Parameter in the Generalized Spline Smoothing Problem},} \textit{The Annals of Statistics}, 13, 1378--1402.

\bibitem[{Yan et~al.(2018)Yan, Paynabar, and Shi}]{tech_2018}
Yan, H., Paynabar, K., and Shi, J. (2018), \enquote{Real-Time Monitoring of High-Dimensional Functional Data Streams via Spatio-Temporal Smooth Sparse Decomposition,} \textit{Technometrics}, 60, 181--197.

\bibitem[{Zhou and Paffenroth(2017)}]{RDA_2017}
Zhou, C. and Paffenroth, R.~C. (2017), \enquote{Anomaly Detection with Robust Deep Autoencoders,} in \textit{Proceedings of the 23rd ACM SIGKDD International Conference on Knowledge Discovery and Data Mining}, ACM, pp. 665--674.

\bibitem[{Zipfel et~al.(2023)Zipfel, Verworner, Fischer, Wieland, Kraus, and Zschech}]{industryAD_2023}
Zipfel, J., Verworner, F., Fischer, M., Wieland, U., Kraus, M., and Zschech, P. (2023), \enquote{Anomaly Detection for Industrial Quality Assurance: A Comparative Evaluation of Unsupervised Deep Learning Models,} \textit{Computers \& Industrial Engineering}, 177, 109045.

\end{thebibliography}
\end{document}